\documentclass[useAMS,usenatbib]{mn2e}
\usepackage{graphicx}

\title[Two-component model of the interaction of an interstellar cloud with surrounding hot plasma]{Two-component model of the interaction of an interstellar cloud with surrounding hot plasma}
\author[E.A. Provornikova et al.]{E.A. Provornikova$^{1,2}$\thanks{E-mail:
provea@iki.rssi.ru (EAP)},V.V. Izmodenov$^{1,2,3}$ and R. Lallement$^{4}$\\
$^{1}$Lomonosov Moscow State University, Moscow, Russia\\
$^{2}$Space Research Institute of RAS, Moscow, Russia\\
$^{3}$Institute for Problems in Mechanics, Moscow, Russia\\
$^{4}$ Service d`Aeronomie du CNRS, Verrieres-le-Buisson, France}
\begin{document}

\date{Accepted yyyy month dd. Received yyyy month dd; in original form yyyy month dd}

\pagerange{\pageref{firstpage}--\pageref{lastpage}} \pubyear{2002}

\maketitle

\label{firstpage}

\begin{abstract}

We present a two-component gasdynamic model of an interstellar cloud embedded in a hot plasma. It is assumed that the cloud consists of atomic hydrogen gas, interstellar plasma is quasineutral. Hydrogen atoms and plasma protons interact through a charge exchange process. Magnetic fields and radiative processes are ignored in the model.  The influence of heat conduction within plasma on the interaction between a cloud and plasma is studied. We consider the extreme case and assume that hot plasma electrons instantly heat the plasma in the interaction region and that plasma flow can be described as isothermal. Using the two-component model of the interaction of cold neutral cloud and hot plasma, we estimate the lifetime of interstellar clouds. We focus on the clouds  typical for the cluster of local interstellar clouds embedded in the hot Local Bubble and give an estimate of the lifetime of the Local interstellar cloud where the Sun currently travels.

The charge transfer between highly charged plasma ions and neutral atoms generates X-ray emission. We assume typical abundance of heavy ions for the Local Bubble plasma and estimate the X-ray emissivity due to charge exchange from the interface between cold neutral cloud and hot plasma. Our results show that charge exchange X-ray emission from the neutral-plasma interfaces can be a non-negligible fraction of the observed X-ray emission.

\end{abstract}


\section{Introduction}

During last decades, after the first discovery of the X-ray emission from comets \citep{b18}, it was found that the X-ray emission produced in charge exchange reactions of highly charged ions with neutrals can be observed almost in every astrophysical object where the mixing of ions and neutrals takes place. In the solar system X-ray emission has been observed from many planets - the Earth, Jupiter, Venus and Mars. \citet{b19} suggested that one of the mechanisms of X-rays from these objects is the charge exchange process, which highly charged ions of the solar wind  undergo in the upper atmosphere of the planets. Also it was found that the charge exchange of solar wind ions and interstellar and geocoronal neutral atoms generates X-ray emission in the heliosphere \citep{b21}. \citet{b1} has pointed out that charge exchange process is the most probable source of X-rays from more distant objects outside of the solar system.
Since the X-ray emission requires the presence of both neutral and ionized gases the following phenomena have been considered as  possible sources of X-rays:  (a) galactic wind interacting with the halo dense cloud; (b) high-velocity cloud moving through halo; (c) dense interstellar clouds moving in intra-cluster gas. Estimates made by \citet{b1}  showed that charge exchange X-ray emissivity from spatially small regions of the neutral-plasma interfaces may have the same order of magnitude as the X-ray emissivity from the hot gas itself.

In order to use X-rays produced by charge transfer for determining the physical properties of plasma-neutral interfaces, these regions need to be studied theoretically. Moreover, the charge exchange changes the structure of the plasma-neutral interfaces. The region where dynamical effects of the charge exchange are intensively explored is the heliospheric interface, i.e. the interface where the solar wind interacts with the local interstellar medium. Due to the charge exchange with both interstellar and solar wind protons the interstellar neutrals provide additional pressure moving the structure closer to the Sun \citep{b22}. The resonant charge exchange is the driver for many physical phenomena in the outer heliosphere and at the heliospheric boundaries \citep{b23}.

Often in exploring the physics of the neutral-plasma interfaces it is necessary to employ a two-component approach since the neutral and charged components are not in equilibrium. Also local equilibrium may or may not occur inside each of the components. If the component is in local equilibrium then it can be described as a fluid, otherwise a kinetic approach is required. For example, a kinetic approach is needed to describe the interstellar neutral component in the heliosphere since the mean free path of the neutrals is comparable with the size of the heliospheric interface region \citep{b20}.

The goal of this paper is to make an initial step in theoretical exploration of the astrophysical phenomena that were selected by \citet{b1} as potential sources of charge-exchange induced X-ray emission. Doing this we develop  a two-component model of cold neutral cloud embedded into a hot plasma. In this initial study we explore the role of proton-neutral charge exchange on the formation and time evolution of neutral-plasma interfaces.

The models of cold cloud - hot plasma interaction have been developed for decades and the main goal of these models was to estimate the lifetime of clouds  (\citet{b6}, \citet{b8}, \citet{b17}, \citet{b9}).  The major difference of our model from the existing models is the two-component approach that we employ as compared with the one-fluid approach used before.

The pioneering theoretical models have considered a cold neutral cloud embedded in a warm gas  (\citet{b4}, \citet{b5})and assumed pressure equilibrium and thermal balance of the cold and warm gases. The model takes into account the thermal conduction and radiative processes of heating and cooling of interstellar medium.  The condition when steady solution exists was found.
This model has been applied for spherical clouds in \citet{b6}. They considered the clouds with the temperature  $T \sim 100$ K and the number density $n \sim 100 \,\ cm^{-3}$. Surrounding warm medium has the temperature $T \sim 10^4$ K and the number density $n \sim 0.1\,\ cm^{-3}$. For these conditions they gave the evaporation rate as a function of a cloud radius and medium pressure which allows to estimate the timescale on which the clouds evaporate. Small clouds with radius $0.02 \leq R_{cloud} \leq 0.03$ pc at the pressure $p/k=10^3\,\ K cm^{-3}$ evaporate in $\sim 10^7$ years.

Later it was found that cold neutral clouds are surrounded by very hot ionized gas. Because of the fact that electrons have high thermal velocities the heat conduction should be taken into account.
 \citet{b8} have considered classical and saturated heat conduction. In absence of magnetic fields and radiative processes the evaporation rate of spherical clouds  was derived for two cases. Subsequent works (\citet{b12}, \citet{b11}, \citet{b10}, \citet{b9}) have extended these models for two and three dimensional geometry and applied magnetic fields, radiative heating and cooling processes and different types of heat conduction.

 The charge exchange process ($H+H^+ \longleftrightarrow  H^+ + H$) was not considered in either of these models. That was because both components have been treated as being in local equilibrium or in a state close to the local equilibrium. In this case the charge exchange does not have any pronounced effect. The situation is different when the components are not in equilibrium as in the case of the heliospheric interface.
 
 When neutral hydrogen gas and hydrogen plasma encounter one another the main process that starts to act is the charge exchange process. This is because the charge-exchange cross section is larger as compared with the corresponding (i.e. momentum transfer) cross section for elastic collisions between the components. The charge exchange is the most effective for momentum exchange between the components. Since the momentum exchange term in the momentum equation for each of the components is proportional to the velocity difference between the components (multiplied by the cross section and the number densities of both components) this term acts as a drag and in fact should reduce the velocity of the neutral component.
 This may results in the increasing of the cloud lifetime as compared with the models where the charge exchange is neglected.
 \par
Another important feature of the model proposed here is the assumption that although the total pressure in the cold cloud is assumed to be in balance with pressure in hot plasma ($P_{cloud} = P_{plasma}$)  there is no pressure balance for each of the components. If we ignore the charge exchange there is nothing that would prevent the neutral cloud from expanding to the ambient plasma and plasma from filling the cloud cavity.  As we will see below the charge exchange is a very efficient process preventing the components from fast penetration to each other.
\par
In Sect. 2 we introduce the two-component model of the interaction of cold neutral cloud and surrounding hot plasma. The problem is formulated in two cases - ``adiabatic" and isothermal. A gasdynamic structure of the plasma-neutral interface is discussed in Sect. 3. Estimates of charge exchange X-ray emission from the interaction interface  are given in Sect.4. The lifetime of the interstellar clouds inside the Local Bubble obtained in the frame of the two-component model is discussed in Sect.5. Conclusions are drawn in Sect.6.

\section[]{ Model}

\subsection{Adiabatic model}

We consider a problem of the interaction of a cold neutral cloud  and hot fully-ionized plasma. The cloud is assumed to be spherically-symmetric and consists of atomic hydrogen gas. Plasma is quasi-neutral and consists of protons, electrons and highly charged ions ($n_e\approx n_p + n_i$). We assume that $n_i/n_p \ll 1$. The medium of hydrogen atoms and ions is described in the two-fluid approach. H atoms and plasma ions interact solely through the charge exchange. In this subsection we  present an ``adiabatic" model of the cloud - plasma interaction. It implies that heat fluxes in the plasma component are ignored, except for the energy transfer between two components due to the charge exchange.

Fluid approach for H atoms is reasonable since (as it is seen below) the frequency of the collisions between H atoms exceeds the frequency of the charge exchange between H atoms and plasma protons  almost in the entire interaction region. This takes place due to relatively large number density of neutrals. Coupling of the plasma and neutral components by the charge exchange results in the momentum and energy exchange between plasma and neutral gas which are not in local thermodynamic equilibrium between each other. In the model we assume that  effective collisions between H atoms lead to local thermodynamic  equilibrium in the neutral component and local thermodynamic equilibrium in the plasma is supported by ``collective" processes.

The governing equations for plasma and hydrogen gas flows are the following:
\begin{eqnarray}
\frac{\partial n_{i}}{\partial t}+\frac{1}{r^2}\frac{\partial r^2 (n u)_{i} }{\partial r}=0,\label{cont} \nonumber\\
\frac{\partial (n u)_{i}}{\partial
t}+\frac{1}{r^2} \frac{\partial\left(r^2 (n u^2+p)_{i}\right)}{\partial r} =\frac {2 p_{i}}{r} +q_{2_{i}}, \label{cont} \nonumber\\
\frac{\partial (n(e+\frac{u^2}{2}))_{i}}{\partial t}+ \frac{1}{r^2}
\frac{\partial\left(r^2(n u(e+\frac{u^2}{2}) +p u)_{i}\right)}{\partial r}
 =q_{3_{i}}, \label{cont} \nonumber
 \end{eqnarray}
where $i=p$ for the plasma and $i=H$ for the hydrogen gas. Here $n_p$ and $n_H$ are the number densities of protons and H atoms, respectively; $e_p=p_p[(\gamma -1)\rho_p]^{-1}$ and $e_H=p_H[(\gamma -1)\rho_H]^{-1}$ are specific internal energies of plasma and hydrogen gas. It is assumed that in the plasma the electron and proton temperatures are equal. The plasma and hydrogen gas temperatures can be determined from the equations of state $p_p=2 n_p k T_p$ and $p_H=n_H k T_H$, respectively.

The terms $q_{2_i}$, $q_{3_i}$ are the sources of momentum and energy due to the charge exchange of H atoms and plasma protons. For the source terms we use the expressions  by \citet{b14}:
$$q_{2_p}=-q_{2_H}=n_p \nu_H (u_H - u_p)$$
$$q_{3_p}=-q_{3_H}=n_p \nu_H \left( \frac{u^2_H - u^2_p}{2} + \frac{2 U^* k}{U m_p}(T_H-T_p) \right)$$
Here $\nu_H$ denotes the charge exchange frequency: $\nu_H = n_H U \sigma^{Hp}_{ex}(U) $, where $\sigma^{Hp}_{ex}(U)$ is the charge exchange cross section and $U$ is the relative velocity of H atom and proton. For charge exchange cross section we use the expression by \citet{b15}: $\sigma^{Hp}_{ex} = (1.64 \cdot 10^{-7} - 6.95 \cdot 10^{-9} ln U)^2$. The expressions for $U^*$ and $U$ can be found in \citet{b14}.
\par 
The initial conditions are the following: $$0<r<R_c: n_H=n_{H1}, T_H=T_1, u_H=0, n_p=0$$
$$ r>R_c: n_p=n_{p2}, T_p=T_2, u_p=0, n_H=0$$
At the initial moment of time a spherical cloud with the radius $R_c$ is filled by the cold hydrogen gas ($T_H \sim 100-10^4$ K) and surrounded by the hot plasma ($T_p \sim 10^6$ K ). It is assumed that at $t=0$ the cold neutral cloud and the hot plasma are in pressure equilibrium: $n_{H1}T_1=2 n_{p2} T_2$. Since $T_1 \ll T_2$ then $n_{H1} \ll n_{p2}$.

We solve the problem in dimensionless form. For the characteristic length the mean free path  of H atom in hot plasma is adopted: $L=[n_0 \sigma^{Hp}_{ex} (u_0)]^{-1}$. For the characteristic velocity it is convenient to take the thermal velocity of hot protons: $u_0=\sqrt{2 k T_2/m_p}$. The number densities are normalized to the proton number density in hot plasma: $n_0=n_{p2}$. Then it is easy to show that the solution depends only on two dimensionless parameters: $\hat{n}=n_{H1}/n_{p2}$ and $\hat{R}_c=R_c/L$.

For the simplicity of the  numerical solution of the problem we introduce small portions of protons and H atoms  in the cloud and hot gas respectively.  $n_{p1}$ is the number density of plasma portion in the neutral cloud and $n_{H2}$ is the number density of H atoms in surrounding plasma. These values are close to zero: $n_{p1} \ll n_{H1}$, $n_{H2} \ll n_{p2}$ and do not affect the solution.

The problem is solved numerically by the Godunov shock capturing method with improved order of scheme accuracy. The space grid has been reconstructed on each time step according to the propagation of a shock wave, a contact discontinuity and a rarefaction wave-front in either neutral or plasma component. To improve the scheme resolution we used the flux limiter function ``min-mod".

\subsection{Isothermal model}

For high temperature plasma the electron heat conduction plays an important role since hot electrons with large thermal velocities  penetrate into the cooler regions of the plasma and heat it by collisions. The thermal conductivity in a fully ionized hydrogen plasma is given by \citep{b7} :$\chi_e = \chi_e (T_p) = a T^{5/2}_p\,\ ergs cm^{-1} s^{-1} K^{-1}$, where $a \sim 10^{-7}\,\ ergs cm^{-1} s^{-1} K^{7/2}$. One can show that since the temperature of the  Local Bubble plasma is about $10^6$ K then the heat flux $q=-\chi_e \partial T_p/\partial r$ is the order of magnitude above the other terms in the energy equation for the plasma. Therefore, we can consider the extreme case, when the electron thermal conduction leads to instant plasma heating in the cloud-plasma interface, and the plasma flow can be described as isothermal.

The system of equations and initial conditions governing the cold cloud interaction with hot isothermal plasma has the same form as for ``adiabatic" model, except the energy equation for the plasma component which is replaced by the condition: $T_p(x,t)=const$.

\section[]{Gasdynamic structure of the neutral cloud-plasma interface}
\subsection{Adiabatic model}

\begin{figure*}
 \vspace*{10pt}
\includegraphics[width = 175 mm]{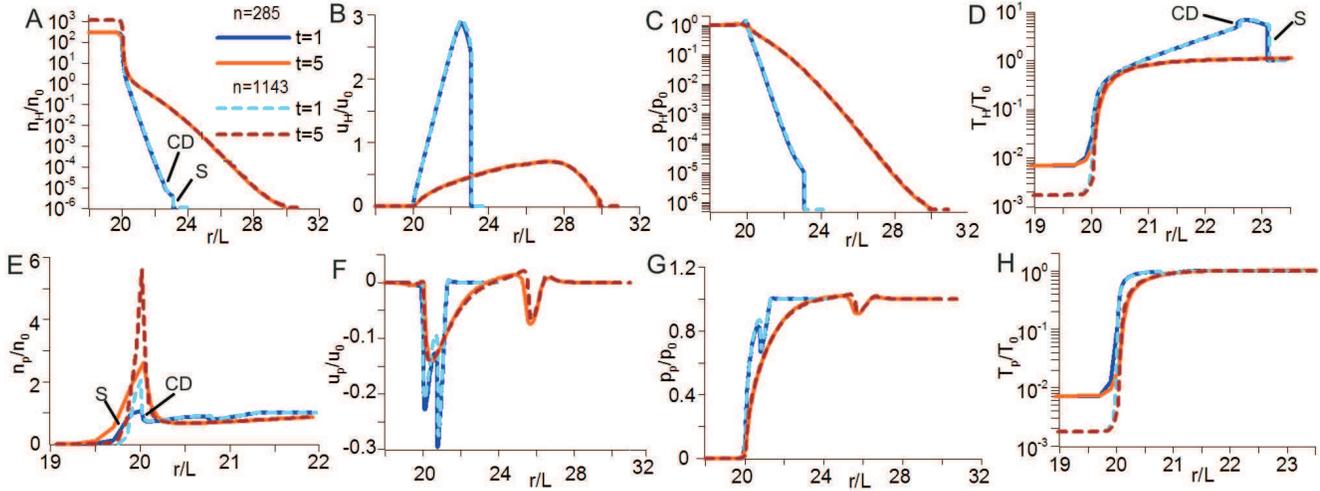}

 \caption{Distributions of dimensionless number density, velocity, pressure and temperature of neutral gas (top plots: A-D) and plasma (bottom plots: E-H) as functions of a radial distance at $\hat{t}=1,5$. Solid curves correspond to the solution for $\hat{n}=285$, dashed curves - $\hat{n}=1143$. $\hat{R}_c=20$. (S) and (CD) denote a shock wave and a contact discontinuity, respectively. }
\label{fig:1}
\end{figure*}

\begin{figure*}
 \vspace*{10pt}
\includegraphics[width = 168 mm]{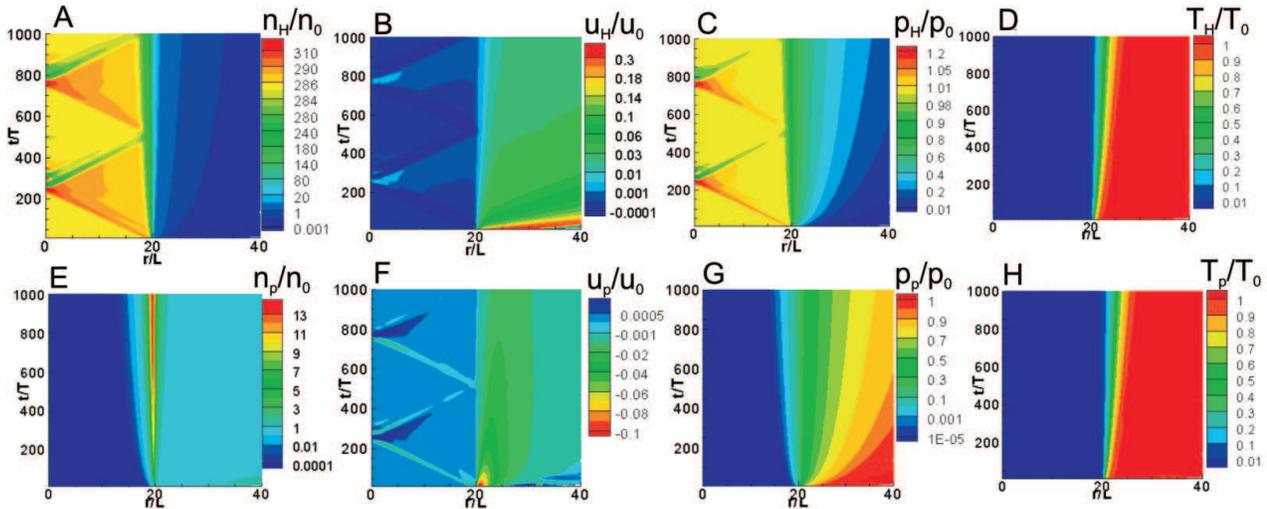}

 \caption{ Space-time distributions of dimensionless number density, velocity, pressure and temperature of neutral gas (top plots: A-D) and plasma (bottom plots: E-H). The solution corresponds to $\hat{n}=285$ and $\hat{R}_c=20$. }
\label{fig:2}
\end{figure*}

In our calculations we varied the dimensionless parameters $\hat{n}$ and $\hat{R}_c$. Here to discuss a qualitative gasdynamic structure of a cloud-plasma interface we present a solution for the values $\hat{n}=285, 1143$ and $\hat{R}_c=20$. It corresponds to all situations when the number densities in cold neutral cloud and surrounding hot plasma satisfy the conditions n$_{H}$=285$ n_{p}$ and n$_{H}$ = 1143$n_{p}$, respectively. The parameter $\hat{n}$ and pressure balance determine the ratio of temperatures in the cloud and the plasma $\frac{T_H}{T_p}$. For example, if we assume $n_p=0.0009\,\ cm^{-3}, T_p=10^6\,\ K $ for the Local Bubble plasma then $\hat{n}=285$  corresponds to the clouds with $n_H=0.25\,\ cm^{-3}, T_H=7000\,\ K$ and $\hat{n}=1143$ - to the clouds with $n_H=1\,\ cm^{-3}, T_H=1750\,\ K$. Radius of the cloud is 2.9 pc.

After the beginning of the interaction the neutral gas flows into the hot plasma and the plasma penetrates into the neutral cloud. The coupling of the plasma and neutral gas in the charge exchange process substantially influences the flows of both components \citep{b24}.  Figure \ref{fig:1} shows number densities, velocities, pressures and temperatures of the hydrogen gas (plots A-D) and the plasma (plots E-H) as functions of a radial distance for two moments of time. After the beginning of the interaction  a contact discontinuity (CD) and a shock wave (S) are formed in the neutral gas and propagate into the hot plasma. It can be seen from plots A-D for $t=1$ in Fig. \ref{fig:1}. The neutral gas is compressed and heated behind the shock. Distributions for $t=5$ show that the charge exchange of H atoms and plasma protons leads to that the shock strength  decreases, and gasdynamic parameters of the neutral gas change continuously.

\begin{figure*}
 \vspace*{10pt}
\includegraphics[width = 175 mm]{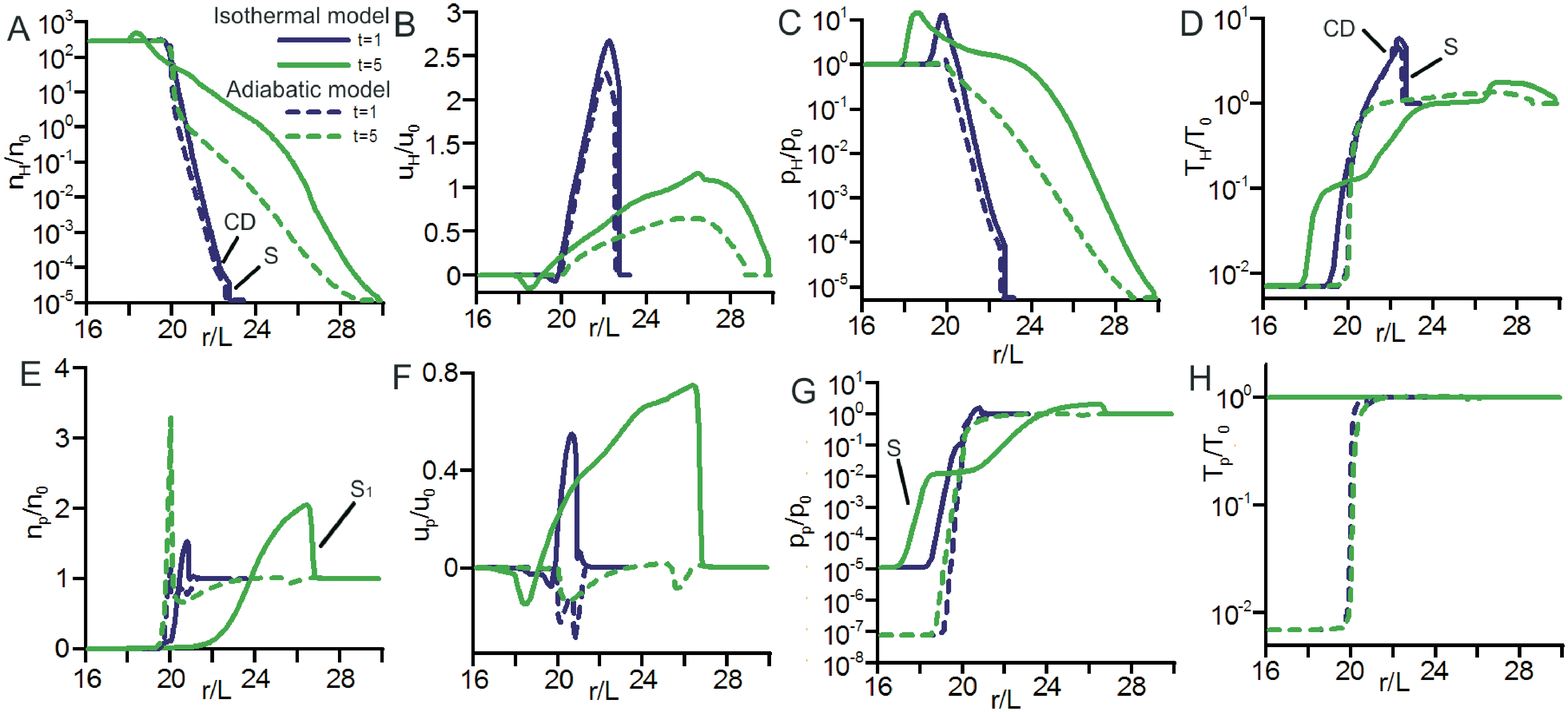}
 \caption{Distributions of gasdynamic parameters of neutral gas (top plots: A-D) and plasma (bottom plots: E-H) as functions of a radial distance for  $\hat{t}=1,5$. Solid curves correspond to the solution in the isothermal model, dashed curves denote the solution in the ``adiabatic" model. Here $\hat{n}=285$ and $\hat{R}_c=20$. (S) and (CD) denote a shock wave and contact discontinuity, respectively. ($S_1$) shows a shock wave formed in plasma.}
\label{fig:3}
\end{figure*}

\begin{figure*}
 \vspace*{10pt}
\includegraphics[width = 150 mm]{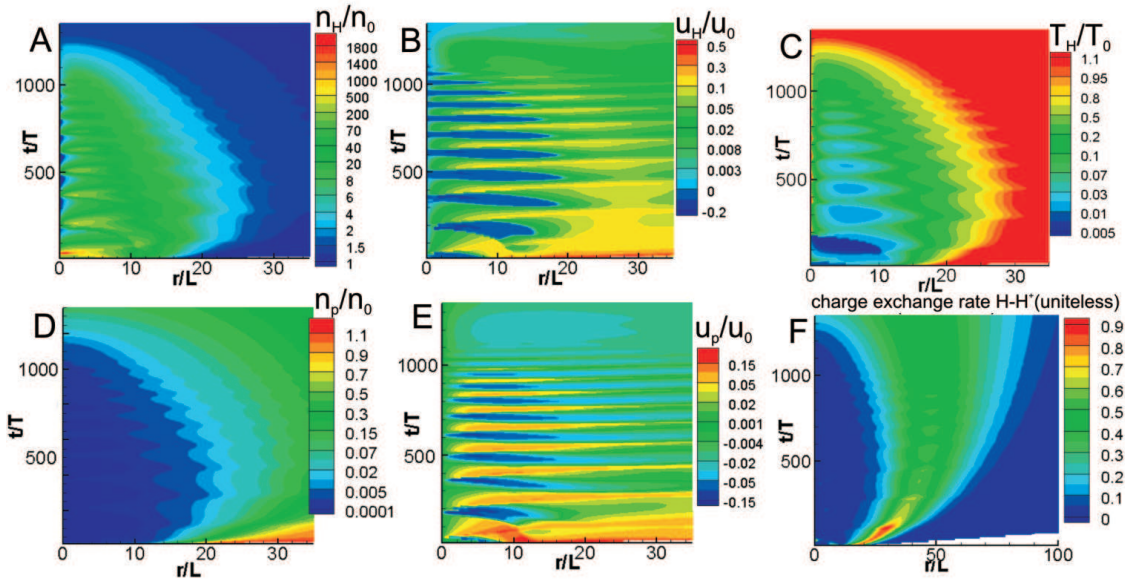}
 \caption{Space-time distributions of number density, velocity and temperature of neutral gas (top plots: A-C) and number density, velocity of plasma (bottom plots: D-E). Plot F shows a charge exchange rate of H atoms and plasma protons. The solution corresponds to $\hat{n}=285$ and $\hat{R}_c=10$. }
\label{fig:4}
\end{figure*}

In the plasma a shock wave (S) and a contact discontinuity (CD) propagate into the neutral cloud at the initial stages of the interaction (Fig. \ref{fig:1}, E). The discontinuities are decelerated very fast due to the interaction with neutral gas by the charge exchange. The plot E in Fig. \ref{fig:1}  shows that at the edge of the cloud the plasma density has a maximum several times larger than the number density in surrounding hot plasma.  Plots B, F in Fig. \ref{fig:1} show that due to the exchange of momentum between the components in the charge exchange process the velocities of neutral gas and plasma decrease with time. It is important to note that at the edge of the  cloud the narrow region is formed where the temperature of both components gradually changes from its value in the cold cloud to the value in the hot plasma. Distributions of hydrogen gas parameters obtained for different values of the parameter $\hat{n}$ show that the solution for the neutral gas beyond the cloud does not depend on the parameter $\hat{n}$. 
\par
Figure \ref{fig:2} presents space-time distributions of the neutral gas and plasma parameters. The results show that the interaction of the neutral gas and hot plasma by charge exchange leads to the formation of the interface region at the edge of the cloud. From the simulation up to  $\hat{t}=1000$, which corresponds to the 1.1 Myr, it is seen that the interface exists during rather long time periods. The neutral cloud-plasma interface is characterized by the following features: 
\par 
1) plasma number density has a maximum at the cloud boundary (Fig.\ref{fig:2}, E). In the interaction region the plasma moves towards the cloud boundary and decelerates due to momentum exchange with H atoms ( Fig. \ref{fig:2}, F). It results in increasing of plasma density with time. For instance, over the interaction period of 1.1 Myr plasma number density increases by 15 times in the transition region.  
\par Since the number density of the hydrogen gas in this region is the same order as the density in the cold cloud (Fig. \ref{fig:2},A) the  frequency of charge exchange between plasma ions and H atoms has a maximum here; 
\par 
2) velocity of neutral gas in plot B of Fig. \ref{fig:2} shows that the cloud does not expand during the interaction with surrounding plasma;
 \par
3) temperature of neutral gas shows that the cloud does not suffer from heating (Fig. \ref{fig:2}, D) and even at large time intervals the gas inside the cloud has an initial temperature $T_H$ . The charge exchange process results in the same temperature of neutral  and plasma components throughout the interface. The temperature increases from cold temperature of the cloud $T_H$ to the high temperature of hot plasma $T_p$ (Fig. \ref{fig:2}, D,H). Numerical solution shows the existence of waves which periodically appear at the edge of the cloud, propagate into the cloud and than reflect back from the cloud center.
\par
Thus in the ``adiabatic" model the cloud remains cold and its boundary is not displaced over a long period of time. It will be seen below that taking into account a heat conduction in the plasma strongly affects this gasdynamic structure of the cloud-plasma interface. But from results of the ``adiabatic" model one can conclude that the charge exchange process may be considered as the most important mechanism providing the existence of the interstellar clouds in the Local Bubble hot plasma. 

\subsection{Isothermal model}

The isothermal model implies that the plasma flow is isothermal everywhere including the interaction region: $T_p(r,t)=T_2$. Figure \ref{fig:3} shows the distributions of gasdynamic parameters for neutral gas and plasma as functions of a radial distance for two moments of time. In fig. \ref{fig:3} we also present the results of the ``adiabatic" model for the sake of comparison.  
\par 
When the cloud gas starts interacting with the ambient plasma the structure of the neutral and plasma flows is similar to the ``adiabatic" case. A contact discontinuity (CD) and a shock (S) are formed in the neutral gas and propagate into the hot plasma (Fig. \ref{fig:3}, A). Since the plasma flow is isothermal only a shock wave exists and tends to move into the cloud (Fig. \ref{fig:3}, G). Plot A in Fig. \ref{fig:3} shows that the number density of H atoms increases at the edge of the cloud and velocity becomes negative(Fig. \ref{fig:3}, B). Since H atoms interact with the hot protons by charge exchange the temperature of H atoms increases in the interface. The narrow region where H atoms are heated and have a maximum density propagates inside the cloud. On the other hand due to the heat flux into the neutral component from hot plasma the pressure of neutral gas increases in the interaction region and gas flows out into the hot plasma. 
\par Momentum exchange of plasma protons and H atoms results in that the plasma velocity becomes positive almost in entire interaction region (comparing to the ``adiabatic" case where in the interface plasma flows towards the cloud with decreasing velocity) (see Fig. \ref{fig:3}, F). Thus the neutral gas expanding to the ambient plasma captures the plasma flow. The region at the edge of the cloud where the plasma density and velocity has a maximum (Fig. \ref{fig:3}, E, F) propagates to the hot medium. Steepening of that region creates a shock wave in the plasma ($S_1$) (Fig. \ref{fig:3}, E).
\par
Comparison of the solutions for the isothermal and ``adiabatic" models in fig.\ref{fig:3} also shows that in the isothermal model disturbances in  neutral gas propagate faster than in the ``adiabatic" model. That occurs due to the heat transfer from hot isothermal plasma into the neutral gas. 
\par
Figure \ref{fig:4} shows space-time distributions of the neutral gas and plasma parameters. In calculations $\hat{n}=285, \hat{R}_c=10$. One can see in figure \ref{fig:4}, plot A that in the isothermal model the size of the cloud changes with time. The cloud expands due to the heat flux from hot isothermal plasma, then cloud radius decreases and the cloud vanishes.

The main results of the modeling of the interaction of cold neutral cloud and hot isothermal plasma are the following:
\par 
1) neutral cloud is being heated with time due to creation of new hot H atoms from hot protons in the charge exchange process. Space-time distributions of neutral gas temperature shows that the temperature of the cloud is equal to the temperature of surrounding plasma at $\hat{t}\approx 1300$ (Fig. \ref{fig:4} C); 
\par 
2) inside the cloud the density of neutral gas decreases with time but plasma density increases (Fig. \ref{fig:4} A,D). Over the period  $\hat{t}\approx 1300$ number density $n_H$ changes from the initial value $\hat{n}_H=285$ to $\hat{n}_H=1$ in the cloud, number density of plasma reaches $\hat{n}_p=0.1$; 
\par 
3) The velocities of  both neutral and plasma components decrease with time due to momentum exchange between the components.

\section{Estimate of charge exchange X-ray emission from cloud-plasma interface}

\begin{figure}
 \vspace{20pt}
\includegraphics[width = 90 mm]{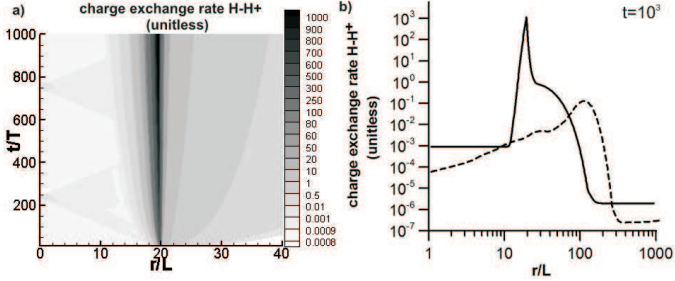}

 \caption{a) Space-time distribution of charge exchange rate $H-H^{+}$ in ``adiabatic" model. b) Charge exchange rate as a function of a radial distance at $\hat{t}=1000$. Charge exchange rate has a maximum in the transition region for both ``adiabatic" (solid curve) and isothermal (dashed curve) models. Here $\hat{n}=285, \hat{R}_c=20$. }
\label{fig:5}
\end{figure}

The distributions of the neutral gas and plasma parameters obtained in the frame of the models allow us to calculate the rate of charge exchange between H atoms and plasma protons $\beta =n_H n_p U \sigma^{Hp}_{ex}$ for the plasma-cloud interaction region. Figure \ref{fig:5} shows a space-time distribution of the charge exchange rate obtained in the ``adiabatic" model. The charge exchange rate has a maximum inside the cloud-plasma interaction interface. This maximum exists due to increased number densities of both plasma and H atoms in this region. This increase of charge-exchange rate is an indication that the charge exchange X-ray emission from the neutral-plasma interface is not negligible as it was pointed out by \citet{b1}. 

\par
 Let us estimate an importance of the charge exchange X-ray emission from the interaction interfaces as compared with the hot gas X-ray emission.
The charge exchange X-ray emissivity is expressed by the integral: $P_{ex}=\int_{r_1}^{r_2} \beta_i dr =\int_{r_1}^{r_2} n_i n_H U \sigma^{Hi}_{ex} dr $, where $r_1$,$r_2$ are boundaries of the transition region, $\beta_i$ is the rate of charge exchange between H atoms and plasma heavy ions, $n_i$ is the number density of heavy ions, $ \sigma^{Hi}_{ex}$ is the charge exchange cross section of H atoms and heavy ions. To get a simple estimate we assume that X-ray photons are generated in the charge exchange process between H atoms and oxygen ions $O^{+7}$ and an abundance of ions in the hot plasma is : $n_{O^{+7}}/n_p \sim 6 \cdot 10^{-5}$.  Here the typical abundance of oxygen ions $O^{+7}$ for solar wind is given. The ratio of charge exchange cross sections is: $\sigma^{HO^{+7}}_{ex}/\sigma^{Hp}_{ex} \sim 1.7 $ \citep{b16}.

\begin{figure}
 \vspace{20pt}
\includegraphics[width = 45 mm]{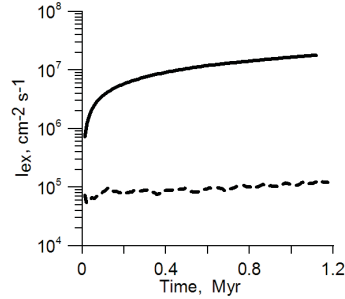}

 \caption{ Time variation of an integral of $H-H^+$ charge exchange rate. The solid curve corresponds to the ``adiabatic" model, dashed curve - isothermal model. $\hat{n}=285.$}

\label{fig:6}
\end{figure}

Therefore $$P_{ex}=\frac{n_{O^{+7}}}{n_p}\frac{\sigma^{HO^{+7}}_{ex}}{\sigma^{Hp}_{ex}}I_{ex},$$ where $I_{ex} = \int_{r_1}^{r_2} n_p n_H U \sigma^{Hp}_{ex} dr$ is the integral of charge exchange rate of H atoms and plasma protons which can be calculated by using our model results. Figure \ref{fig:6} shows the dependence of $I_{ex}$ on time.  In the ``adiabatic" case for the Local interstellar cloud with parameters $n_H=0.25\,\ cm^{-3}, T_H=7000\,\ K$ corresponding to dimensionless parameter $\hat{n}=285$ one can estimate:$$P_{ex}\approx 10^3 keV cm^{-2} s^{-1}.$$ The volume emissivity of diffuse X-ray emission from hot gas can be estimated as: $P_h=5.8 \cdot 10^{-14} n^2_e keV cm^{-3} s^{-1},$ where $n_e$ is the number density of electrons in hot plasma \citep{b1}. In our calculations $n_e=0.0009\,\ cm^{-3}$. Then $P_h\approx 4\cdot 10^{-20} keV cm^{-3} s^{-1}$. Assuming the characteristic size of the Local Bubble is $L_{LB} \approx$ 100 pc the emissivity of diffuse X-ray from the hot plasma is $P_h L_{LB} \approx 10 keV cm^{-2} s^{-1}$. The emissivity ratio is: $P_{ex}/P_h L_{LB} \approx 10^2$. The charge exchange X-ray emissivity from the cloud-plasma interface is two orders above the  thermal emissivity from hot gas. Since ``adiabatic" model does not take into account thermal conduction processes the calculated magnitude of charge exchange X-ray emissivity is overestimated.

Results of the isothermal model also show that the charge exchange rate has a maximum in the neutral cloud - plasma interface (Fig. \ref{fig:4} F), but the value of the maximum is much less than in the ``adiabatic" model (Fig. \ref{fig:5}, b). Using the average value of $I_{ex}$ (see fig. \ref{fig:6}, dashed curve) for the Local interstellar cloud the charge exchange X-ray emissivity is estimated as: $P_{ex} \approx 10\,\ keV cm^{-2} s^{-1}$. Using the estimate for the diffuse  hot plasma X-ray emissivity from the Local Bubble one can see the ratio $P_{ex}/P_h L_{LB} \approx 1$. Thus even in the isothermal case (which provides us low limit of X-rays) the emissivity from the neutral-plasma interface due to charge exchange has the same order of magnitude as compared with the hot gas X-ray emissivity. 

\section{Lifetime of the interstellar clouds}

Let us define a lifetime of a cloud as a timescale over which the temperature in the cloud becomes equal to the temperature in surrounding hot medium. Our two-component model allows to give an estimate of the lifetimes of  interstellar clouds surrounded by hot Local Bubble plasma. This can be done for  two extreme cases - the ``adiabatic" and isothermal. 
\par In section 2 we showed that in the frame of the ``adiabatic" model the cloud boundary is nearly steady and the cloud may exist more than several Myr. For example, simulation of the Local interstellar cloud (LIC) with parameters $n_H =0.25 \,\ cm^{-3}, T_H=7000\,\ K$ embedded in the Local Bubble with $n_p=0.0009 \,\ cm^{-3}, T_p=10^6\,\ K$ shows that the LIC lifetime may reach more than $5.6$ Myr.This estimate changes essentially when an electron heat conduction is taken into account. \par
\begin{figure}
 \vspace{10pt}
\includegraphics[width = 70 mm]{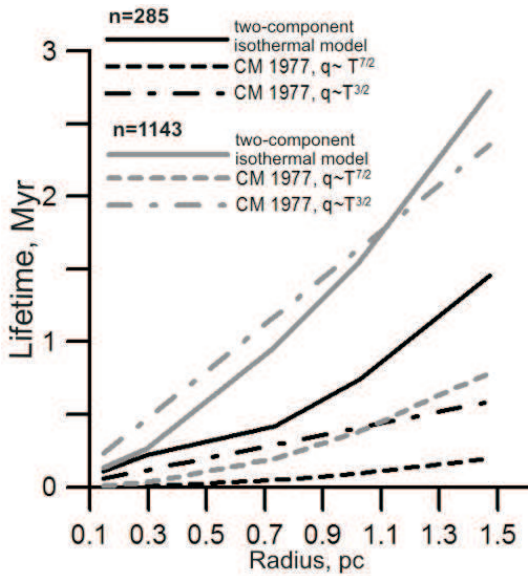}

 \caption{ The lifetime of interstellar clouds as a function of cloud radius. Black curves correspond to the solution for $\hat{n}=285$, grey curves - $\hat{n}=1143$. Solid curves denote the results of two-component isothermal model, dashed curves - \citet{b8} in the case of classical thermal conduction, dash-dot curves - \citet{b8} in the case of saturated heat conduction.}

\label{fig:7}
\end{figure}

 We performed model calculations for different radii of the cloud. Figure \ref{fig:7} shows the dependence of the cloud lifetime on the cloud radius. The results are presented for two values of the dimensionless parameter $\hat{n}=285,1143$. Solid curves denote the results of two-component isothermal model. The lifetime of the  clouds with the number density $\sim 0.25\,\ cm^{-3}$ and temperature $\sim 7000$ K is estimated as 0.2-1.5 Myr. The colder clouds with the number density $\sim 1\,\ cm^{-3}$ and  temperature $\sim 1700$ K may exist twice as long: the lifetime of the clouds with the radius $\sim$ 1.1-1.5 pc are $\sim$ 1.6-2.7 Myr. It is known that the Local interstellar cloud has irregular shape with the characteristic scale about 1-10 pc. If we adopt the size of the LIC as 3 pc then its lifetime is estimated as $\sim$ 1.5 Myr. \par
 
In \citet{b8} the analytical formulae for the lifetime of the spherical neutral clouds embedded in  hot plasma were derived. They considered two types of heat conduction in fully ionized hydrogen plasma: 1) classical thermal conduction $q \sim T^{7/2}$; 2) saturated heat conduction $q\sim T^{3/2}$, when a mean free path of electrons in plasma is comparable to the temperature scale height.
Magnetic fields and radiation processes were ignored in the model. Dashed curves in fig. \ref{fig:7} show the lifetime calculated in the case 1). Dash-dotted curves denote the results obtained in the case 2). For the clouds with the radii greater than $1.1$ pc the lifetime obtained in the two-component model exceeds the lifetime calculated in \citet{b8} in both cases 1) and 2). For denser clouds ($\hat{n}=1143$) with radii $0.1-1.1$ pc the lifetime in the two-component model is comparable with the lifetime in the case 2) with the saturated heat conduction. Note, that our two-component model takes into account the charge exchange process and the plasma heating by the heat conduction. The latter process is considered in extreme case of the isothermal plasma flow. For more realistic situation the effect of heat conduction is expected to be somewhat smaller and, therefore, the cloud lifetime will be larger. Nevertheless, the comparison of lifetime estimates with the models \citet{b8} demonstrates that the lifetime in our model is larger or comparable with \citet{b8}. That means that the charge exchange process plays a significant role in the evolution of the interstellar clouds and should be taken into account.

\section{Conclusions}
We have presented a two-component gasdynamic model of the interaction of a cold neutral cloud and surrounding hot plasma. It was assumed that the neutral and plasma components interact by the charge exchange process. The problem has been considered in two extreme cases - ``adiabatic" and isothermal. In the ``adiabatic" model the heat conduction in plasma is ignored. In the isothermal model it is assumed that the plasma is isothermal due to effective electron heat conduction. We obtained gasdynamic solutions which describe the structure of the neutral-plasma interfaces.
   The results of the ``adiabatic" model can be briefly summarized as follows:
\begin{itemize}
\item The interaction of cloud neutral gas and plasma through charge exchange process results in the formation of a plasma-neutral interface, i.e. the transition region at the edge of the cloud;
\item In the transition region the number densities of both neutral and plasma components increase significantly and become non-negligible;
\item Inside the transition region the temperature of the medium (for both plasma and neutral) has all of the intermediate values;
\item Over the entire period of the cloud-plasma interaction the cloud remains cold  and its boundary is nearly steady;
\item The emissivity of X-ray generated by the charge exchange process of plasma heavy ions and H atoms in the neutral-plasma interface is two orders of magnitude larger than the emissivity of diffuse X-ray from the hot medium. 
\end{itemize}

The basic results obtained in the isothermal model are the following:
\begin{itemize}
\item The interaction of the neutral cloud with hot isothermal plasma results in heating and disappearing of the cloud;
\item  The lifetime of the interstellar clouds with the number density $\sim 0.25\,\ cm^{-3}$ and temperature $\sim 7000$ K surrounded by the Local Bubble plasma with $n_p=0.0009\,\ cm^{-3}$ and $T_p=10^6$ K is estimated as $0.2-1.5$ Myr. The lifetime of the Local interstellar cloud is estimated as $1.5$ Myr;
\item We conclude also that the charge-exchange X-ray emissivity from the neutral-plasma interfaces may have the same order of magnitude as diffuse X-ray emissivity from the extended hot plasma regions. It means that in the analysis of experimental data on X-ray emission in different astrophysical cases where the neutral or partly ionized gas interacts with plasma it is important to consider the contribution of X-rays induced by the charge exchange.
\end{itemize}

\section*{Acknowledgments}
This work was supported by RFBR grant 10-02-93113. E.A.P. is also supported by Russian Federal Special-Purpose Program (state contract No 16.740.11.0309).

\bsp

\label{lastpage}

\end{document}